\documentclass[fp,twocolumn]{jpsj3}
\usepackage{txfonts}
\usepackage{mathtools,nccmath}
\usepackage{lipsum}
\usepackage{cuted}
\usepackage{amsfonts,amssymb,enumerate, epstopdf,xcolor,mathptmx}

\DeclarePairedDelimiterX{\bra}[1]{\langle}{\rvert}{#1\,}
\DeclarePairedDelimiterX{\ket}[1]{\lvert}{\rangle}{\,#1}
\DeclarePairedDelimiterX{\makebraket}[1]{\langle}{\rangle}{#1}

\NewDocumentCommand{\braket}{som}{%
  \begingroup\activatebraketbar
  \IfBooleanTF{#1}
    {\makebraket*{#3}}
    {\IfNoValueTF{#2}{\makebraket{#3}}{\makebraket[#2]{#3}}}%
  \endgroup
}

\makeatletter
\newcommand{\braketbar}{%
  \,\delimsize\vert\@ifnextchar|{\!}{\,}%
}
\makeatother
\newcommand{\activatebraketbar}{%
  \begingroup\lccode`~=`|\lowercase{\endgroup\let~}\braketbar
  \mathcode`|="8000
}

\newcommand{\be}{\begin {equation}} \newcommand{\ee}{\end {equation}}
\newcommand{\beqa}{\begin {eqnarray}} \newcommand{\eeqa}{\end {eqnarray}}

\newcommand{\mbf}{\textbf}

\title{Electron excitation rate in dielectrics under an intense elliptically polarized laser field}

\author{Prachi Venkat\thanks{venkat.prachi@qst.go.jp} and Tomohito Otobe\thanks{otobe.tomohito@qst.go.jp}}
\inst{ Kansai Photon Science Institute, National Institutes for Quantum  Science and Technology (QST), Kizugawa, Kyoto 619-0215, Japan \\}

\abst{Electron excitation in dielectrics is studied for an elliptically polarized laser field. As the first step, we develop an analytical 
formula for electron excitation rate under elliptically polarized laser as the extension of our previous work [T. Otobe \textit{et. al.}, JPSJ \textbf{88} (2019) 024709].
In the next step, we calculate the excitation rate depending on the band structure by assuming direction dependence of reduced mass.
We find that although the ellipticity decreases the excitation rate significantly in an isotropic system, the energy oscillation of  electrons 
due to the intra-band dynamics in the anisotropic band structure increases the excitation rate with higher ellipticity. 
Our results indicate that we can control the excitation rate in dielectrics by varying the ellipticity, depending on the band anisotropy.
}

\begin{document}
\maketitle
\section{Introduction}
Electron excitation in dielectrics by an intense laser field is
the main process in laser-matter interactions. Technical
developments in femtosecond laser processing have made it
possible to produce a nano-scale laser-induced periodic
surface structure (LIPSS)\cite{Davis:96,Miura97,Corkum-06,Gattass:2008aa}, and to realize non-thermal ablation
for sub-wavelength resolution. In particular, for femtosecond
lasers, electron excitation by multi-photon absorption
and tunneling process is crucial because such non-linear
processes generate a controllable free-carrier density and
confine material change to the focal volume. Therefore, the
prediction of electron excitation rate using theoretical models
and/or numerical simulation is important.
 Non-linear carrier dynamics finds application in various fields of laser-matter interaction such as HHG \cite{Ghimire-2011, You-2017}, transient transmission changes the carrier-envelope-phase-dependent electronic currents \cite{Schiffrin-2013}.

Electron excitation can be carried out by various mechanisms such as multi-photon absorption and tunneling, depending on the laser parameters. Various models have been proposed over the years to model multi-photon processes and tunneling of electrons numerically. The Keldysh theory describes electron excitation in atoms and crystals \cite{Keldysh-1965,Golin-2014}, and various mechanisms have been developed based on the Keldysh approach \cite{Faisal-1973, Jones-77, Reiss-1980, ADK-1986, Gruzdev-07,Otobe-2019}. 

Incident laser intensity is a paramount factor in deciding whether the interaction follows linear or non-linear dynamics. Similarly, the polarization of the incident laser field significantly affects the carrier excitation process and optical properties of a material. The time-resolved dynamic Franz--Keldysh effect, which is responsible for the change in the dielectric function of target material, depends significantly on the polarization of incident field \cite{Otobe-dfke-2016}. 

Change in the polarization of laser field also affects the dynamic phase of electrons undergoing intra-band motion, ultimately affecting the excitation rate during multi--photon absorption.
For example, circularly polarized lasers are also important as an
ultrafast laser waveguides\cite{Nejadmalayeri:06}, and in controlling laser--induced nonastructure \cite{Huang:08}.
M. Koz\'ak \textit{et. al.} have reported that 
the ellipticity of the laser polarization decreases the excitation rate in diamond \cite{Kozak-2019}. 
On the other hand, V. V. Temnov \textit{et. al.} have reported that the electron excitation rate induced by a circularly polarized laser is twice that induced by a linearly polarized laser at the same laser irradiance in fused silica and sapphire \cite{Temnov-06}.
We have to study the laser-matter interaction with elliptically polarized laser
in order to understand the above two conflicting experiments.

The electron excitation dynamics has been numerically studied, employing the time-dependent density functional theory (TDDFT) \cite{Otobe-2008,Shinohara-2010,Yabana-2012} as well as analytical models \cite{Otobe-2017,Otobe-2019}.
Although the numerical simulation employing TDDFT is one of the most accurate approaches, it needs a sizeable computational resource, and analyzing the details of the physical process is sometimes demanding.
The analytical approach is not accurate like TDDFT, 
although we can use it to study the physics of the interaction in detail.

In this paper, we adopt an analytical approach to obtain the electron excitation rate for the interaction of isotropic dielectrics with an intense elliptically polarized laser field. The formulation is based on the general formula presented in the previous study \cite{Otobe-2019}.  
The effect of ellipticity of incident laser field on excitation rate is also studied. Furthermore, we study the dominance of $n$-order photonic processes at different intensities for elliptical polarization. 
We also extended the fundamental equation to the anisotropic band by assuming the 
direction-dependent reduced mass. We performed real-time simulations for 
carrier density after the laser pulse ends and analyzed the effect of 
anisotropy of the system.

This paper is organized as follows: Section \ref{form} describes the analytical formulation of the excitation rate for elliptically polarized field with isotropic and anisotropic band structures, section \ref{results} presents results of electron excitation in a diamond for different laser and material parameters, and then finally, we would like to summarize the work in section \ref{conclusion}.

\section{Formulation}
\label{form}

Starting with the Schr\"odinger equation which is assumed as, 
\begin{equation} 
\epsilon^{G}_{n,\mbf{k}}u^{G}_{n,\mbf{k}}(\mbf{r}) = \left( \frac{1}{2m}(\mbf{p} + \hbar\mbf{k})^2 + V(\mbf{r}) \right)u^{G}_{n,\mbf{k}}(\mbf{r}) 
\end{equation}
$\mbf{p}$ being the momentum operator and $\mbf{k}$ being the Bloch wave vector. $V(\mbf{r})$ is the potential and $\epsilon^{G}_{n,\mbf{k}}$ and $u^{G}_{n,\mbf{k}}(\mbf{r})$ are the energy and wave functions, respectively, corresponding to the $\mbf{k}$ and $n^{th}$ band. Both $V(\mbf{r})$ and $u^{G}_{n,\mbf{k}}(\mbf{r})$  are periodic in space.
The time-dependent Schr\"odinger equation is described as \begin{equation} 
\left[ \frac{1}{2m}\left( \mbf{p} + \hbar\mbf{k} + \frac{e}{\hbar c}\mbf{A}(t) \right)^2 + V(\mbf{r}) \right]u_{n,\mbf{k}}(\mbf{r},t)=i\hbar\frac{\partial}{\partial t}u_{n,\mbf{k}}(\mbf{r},t), 
\end{equation}
where $\mbf{A}(t)$ is the vector potential for an elliptically polarized field \cite{Otobe-2016}:
\begin{equation}
\vec{A}(t) = A_0(\eta \sin(\omega t),0,\cos(\omega t)).
\end{equation}
Here, $\eta$ is the parameter that decides the ellipticity of the field and is $\eta=0(1)$ for linear(circular) polarization.
Varying the $\eta$ between $0$ and $1$ will change the ellipticity of  polarization. 

The time-dependent wave function, $u_{n,\mbf{k}}(\mbf{r},t)$, can be expressed by Houston function, 
\begin{equation} 
w_{n,\mbf{k}}(\mbf{r},t) = u^{G}_{n,\mbf{k}+\frac{e}{c}\mbf{A}(t)}(\mbf{r})\exp\left[ -i\int^{t}\epsilon^{G}_{n,\mbf{k}+\frac{e}{c}\mbf{A}(t')}dt' \right], 
\end{equation}
as
\begin{equation}
u_{n,\mbf{k}}(\mbf{r},t)=\sum_{n'}C_{nn'}^{\mbf{k}}(t)w_{n',\mbf{k}}(\mbf{r},t).
\end{equation}
Here, $u^{G}_{n,\mbf{k}}(\mbf{r})$ is the wave function of ground state.
As presented in Ref.\cite{Otobe-2019}, we can obtain a simplified expression for the time-evolution of the  coefficient $C_{nn'}^{\mbf{k}}(t)$ as follows:
\begin{equation} 
\label{eq:td_C}
\begin{split} 
i\frac{\partial C^{\mbf{k}}_{nn'}(t)}{\partial t} = & i\frac{e}{mc}\frac{d\mbf{A}(t)}{dt}\sum_{n''}C^{\mbf{k}}_{nn''}(t)\braket*{ u^{G}_{n',\mbf{k}+\frac{e}{c}\mbf{A}(t)}|\frac{\partial u^{G}_{n'',\mbf{k}+\frac{e}{c}\mbf{A}(t)}}{\partial \mbf{k}}}\\
& \times \exp\left[ -i\int^{t}dt'\Delta\epsilon^{G}_{n"n',\mbf{k}+\frac{e}{\hbar c}\mbf{A}(t')} \right]. 
\end{split}
\end{equation}
The important quantity in Eq.~(\ref{eq:td_C}) is the time integration of the relative energy, $\Delta\epsilon^{G}_{n"n',\mbf{k}+\frac{e}{\hbar c}\mbf{A}(t)}=\epsilon^{G}_{n'',\mbf{k}+\frac{e}{\hbar c}\mbf{A}(t)} -\epsilon^{G}_{n',\mbf{k}+\frac{e}{\hbar c}\mbf{A}(t)}$, because it includes multi-photon absorption processes as discussed in following sections.

\subsection{Isotropic band}
In the following discussion, we assume the isotropic parabolic two-band system,
\begin{equation} \label{band} 
\epsilon^{G}_{c,\mbf{k}} - \epsilon^{G}_{v,\mbf{k}} =\Delta\epsilon^{G}_{cv,\mbf{k}} =E_g + \frac{\hbar^2\mbf{k}^2}{2\mu},
\end{equation}
to simplify the equations and results.
Here $E_g$ is the band gap, $\mu$ is the reduced mass and the subscript $c$ ($v$) corresponds to conduction (valence) band.

If we assume the transition dipole momentum to be isotropic, the transition coefficient for this model is simplified to 
\begin{equation} \label{coeff}
\begin{split}
C^{\mbf{k}}_{vc}= &-\frac{ieP_{cv}A_0}{2\hbar mc}\int_{-T}^{T}dt[M^{-}e^{i\omega t} + M^{+}e^{-i\omega t}]\\
&\times \exp\left[ \frac{i}{\hbar}\int^{t}\Delta\epsilon^{G}_{cv,\mbf{k}+\frac{e}{\hbar c}\mbf{A}(t')}dt' \right]
\end{split}
\end{equation}
where $M^{\pm}=1\pm i\eta$, and $P_{cv}$ is the absolute value of the transition dipole momentum, it can be estimated as $P_{cv}\simeq m\sqrt{E_g/4\mu})$ \cite{Kane-1960}. The exponential term in Eq.~(\ref{coeff}) can be expressed as, 
\begin{equation}
\begin{split}
 &\exp\left[\frac{i}{\hbar}\int^{t}\Delta\epsilon_{cv,\mbf{k}+\frac{e}{\hbar c}\mbf{A}(t')}^{G}dt'\right] \\
 &=\sum_{l,m,n} \exp \left[ \frac{i}{\hbar}\left( E_g + \frac{\hbar^2 k_0^2}{2\mu} + \frac{e^2A_0^{2}}{4\mu c^2}(1+\eta^2) \right)t  \right] \\ 
 & \times J_{l}(\alpha)
 J_{m}(\beta)
 J_{n}(\gamma)e^{i(2l+m+n)\omega t}e^{i m\frac{\pi}{2}}
\end{split}
\end{equation}
where $\theta$ and $\phi$ are the angle of $\mbf{k}$  
from $z$- and $x$-axis respectively, 
and $J_i$ is the $i$-th order 
Bessel function with arguments:
\begin{equation}
\alpha = (1-\eta^2)\frac{e^2A_0^{2}}{8\hbar\mu\omega c^2}
\end{equation}
\begin{equation}
\beta = \eta\frac{ek_0A_0\sin\theta\cos\phi}{\mu c\omega }
\end{equation}
\begin{equation}
\gamma = \frac{ek_0A_0\sin\theta}{\mu c \omega}.
\end{equation}

 The transition rate for every $\mbf{k}$ is $w_{\mbf{k}} = \lim_{T\to\infty}\frac{|C_{cv}^{\mbf{k}}|^2}{2T}$.
The total transition rate per unit time ($W$) is obtained by 
\begin{equation}
\begin{aligned}
\begin{split}
W = & \sum_{k}w_{\mbf{k}}=\int\frac{2k_0^2}{8\pi^3}dk_0\sin\theta d\theta d\phi w_{\mbf{k}}
\end{split}
\end{aligned}
\end{equation}
 The final expression for the transition rate is 
\begin{equation}
\label{eq:W_fin}
\begin{aligned}
\begin{split}
W &= \frac{e^2P_{cv}^2A_{0}^{2}\mu^{3/2}}{4\sqrt{2}\pi^2\hbar^{4}m^{2}c^{2}}
\sum_{L=l_0}^{\infty}\sum_{m_1,m_2}\int d\theta \sin\theta\int d\phi\\ &(-1)^{m_1-m_2}\sqrt{\kappa_L}\Big[(1+\eta^2)\Big\{\\
&J_{L-m_1-1}(\tilde{\gamma},\tilde{\alpha})J_{m_1}(\tilde{\beta})J_{L+m_1-2m_2-1}(\tilde{\gamma},\tilde{\alpha})J_{2m_2-m_1}(\tilde{\beta})\\
&+J_{L-m_1+1}(\tilde{\gamma},\tilde{\alpha})J_{m_1}(\tilde{\beta})J_{L+m-2n+1}(\tilde{\gamma},\tilde{\alpha})J_{2m_2-m_1}(\tilde{\beta})\Big\} \\
+&2(1-\eta^2)J_{L-m_1-1}(\tilde{\gamma},\tilde{\alpha})J_{m_1}(\tilde{\beta})\\
&\times J_{L+m_1-2m_2+1}(\tilde{\gamma},\tilde{\alpha})J_{2m_2-m_1}(\tilde{\beta})\Big],
\end{split}
\end{aligned}
\end{equation}
where $J_i(a,b)$ is the $i$-th order of generalized Bessel function \cite{Reiss-2003}, $L$ is the order of multi-photon absorption, 
 $l_0$ is the lowest photon process order to overcome the 
band gap with ponderomotive energy
\begin{equation}
    \kappa_L =L\hbar \omega-\left\{E_g+(1+\eta^2)\frac{e^2A_0^2}{4\mu c^2}\right\}\ge 0,
\end{equation}
and the new arguments for Bessel functions
($\tilde{\alpha},\tilde{\beta},\tilde{\gamma}$) are 
\begin{equation}
\tilde{\alpha} = (\eta^2-1)\frac{e^2A_0^{2}}{8\hbar\mu\omega c^2}
\end{equation}
\begin{equation}
\tilde{\beta} = \eta\frac{e\sqrt{2\kappa_L}A_0\sin\theta\cos\phi}{\sqrt{\mu} c\omega }
\end{equation}
\begin{equation}
\tilde{\gamma} = \frac{e\sqrt{2\kappa_L}A_0\sin\theta}{\sqrt{\mu} c \omega}.
\end{equation}
To derive Eq.~(\ref{eq:W_fin}), we impose symmetry in the integration about $\phi$. 
\subsubsection{linear and circular polarization limit}
From the definition of the ellipticity $\eta$, 
we can reproduce the linear and circular polarization cases
as the limit of $\eta=0$ and 1 of Eq.~(\ref{eq:W_fin}).
For $\eta=0$, the excitation rate with linearly polarized laser is
\begin{equation}
\label{eq:Lin}
\begin{aligned}
\begin{split}
    W =& \frac{e^2P_{cv}^2A_{0}^{2}\mu^{3/2}}{2\sqrt{2}\pi\hbar^{4}m^{2}c^{2}}\\
\times&\sum_{L=l_0}^{\infty}\int d\theta \sin\theta  \sqrt{\kappa_L} \left(J_{L+1}(\tilde{\gamma},\tilde{\alpha})+J_{L-1}(\tilde{\gamma},\tilde{\alpha})\right)^2
\end{split}
\end{aligned}
\end{equation}
with $\tilde{\alpha}=-e^2A_0^{2}/8\hbar\mu\omega c^2$ and
$\kappa_L=L\hbar \omega-\left(E_g+e^2A_0^2/4\mu c^2\right)$, while  $\eta=1$ gives the rate with circularly polarized laser as
\begin{equation}
\label{eq:Circ}
\begin{aligned}
\begin{split}
W =& \frac{e^2P_{cv}^2A_{0}^{2}\mu^{3/2}}{2\sqrt{2}\pi^2\hbar^{4}m^{2}c^{2}}
\sum_{L=l_0}^{\infty}\sum_{m_1,m_2}\int d\theta \sin\theta\int d\phi (-1)^{m_1-m_2}\sqrt{\kappa_L}\\ &\times \Big\{J_{L-m_1-1}(\tilde{\gamma})J_{m_1}(\tilde{\beta})J_{L+m_1-2m_2-1}(\tilde{\gamma})J_{2m_2-m_1}(\tilde{\beta})\\
&+J_{L-m_1+1}(\tilde{\gamma})J_{m_1}(\tilde{\beta})J_{L+m_1-2m_2+1}(\tilde{\gamma})J_{2m_2-m_1}(\tilde{\beta})\Big\},
\end{split}
\end{aligned}
\end{equation}
with $\tilde{\beta}=(ek_0A_0/\mu c\omega) \sin\theta\cos\phi$ and 
$\kappa_L=L\hbar \omega-\left(E_g+e^2A_0^2/2\mu c^2\right)$.
To derive Eq.~(\ref{eq:Circ}), we use the relation of 
 generalized Bessel function $J_i(a,0)=J_i(a)$.
Although Eq.~(\ref{eq:Lin}) is consistent with our previous result \cite{Otobe-2019}, Eq.~(\ref{eq:Circ}) is not because the definition
of polarization direction is different.

\subsection{Band structure dependence}
In general, the reduced mass $\mu$ around the band-edge has direction dependence
due to the band structure, $\mu\rightarrow \mu(\theta,\phi)$.
Under the vector potential $\mbf{A}(t)$, the momentum of electron in 
the Bloch phase space is changed from $\mbf{k}=k_0(\sin\theta \cos\phi, \sin\theta \sin\phi, \cos\theta)$ to 
\begin{equation*}
\begin{aligned}
\begin{split}
    \mbf{k'}(t)&=(k_0\sin\theta \cos\phi +\frac{e}{\hbar c}A_0\eta \sin\omega t, \\
    &k_0\sin\theta \sin\phi, \\
    &k_0\cos\theta+\frac{e}{\hbar c}A_0\cos\omega t)\\
    &\equiv(k'_0\sin\theta' \cos\phi', k'_0\sin\theta' \sin\phi', k'_0\cos\theta').
    \end{split}
    \end{aligned}
\end{equation*}
We have a new time-dependent Bloch wavevector with, 
\begin{equation}
\begin{aligned}
\begin{split}
    k'_0(t)&=\Big(k_0^2+\frac{e^2}{\hbar^2c^2}A_0^2(\cos^2\omega t+\eta^2\sin^2\omega t)\\
   &+2\frac{e}{\hbar c}k_0A_0(\cos\theta\cos\omega t+\eta\sin\theta\cos\phi\sin\omega t)\Big)^{1/2}
    \end{split}
\end{aligned}
\end{equation}
\begin{equation}
\theta'(t)=\arccos\left[\frac{k_0\cos\theta+\frac{e}{\hbar c}A_0\cos\omega t}{k'_0(t)}\right]
\end{equation}
, and 
\begin{equation}
\phi'(t)=\arctan\left[\frac{k_0\sin\theta\sin\phi}{k_0\sin\theta\cos\phi+\frac{e}{\hbar c}A_0\eta\sin\omega t}\right], 
\end{equation}
we can calculate
the energy phase including band structure as
\begin{equation}
\label{newKph}
\begin{split}
 &\exp\Big[\frac{i}{\hbar}\int^{t}\Delta\epsilon^{G}_{cv,\mbf{k}+\frac{e}{\hbar c}\mbf{A}(t')}dt'\Big] \\
 &=\exp\Big[\frac{i}{\hbar}\int^{t} \left(B_g+\frac{\hbar^2\mbf{k}'^2_0}{2\mu(\theta',\phi')^2}\right)dt'\Big].
\end{split}
\end{equation}
For the case of the simple $M$-fold symmetric system in $\theta$ and $\phi$, it can be expressed as
\begin{equation}
\label{mu_t_p}
    \mu(\theta,\phi)=\mu_0\left[1+\xi \left(\sin^2 \frac{M}{2}\theta +\sin^2 \frac{M}{2}\phi\right) \right].
\end{equation}
The relative energy and 
transition dipole moment have the $M$-fold symmetry with the reduced mass in Eq.~(\ref{mu_t_p}).
By substituting the  reduced mass to the general formula of the transition coefficient Eq~(\ref{eq:td_C}), we can get the transition coefficient with band structure,
\begin{equation}
\label{TD_Ccv}
    C^{\mbf{k}}_{vc}(t)=-e\int^t dt'\frac{\mbf{E}(t')\cdot\mbf{P}_{\mbf{k'}(t'),cv}}{E_g+\frac{\hbar\mbf{k'}(t')^2}{2\mu(\theta',\phi')}}\exp\left[\frac{i}{\hbar} \int^{t'}dt" \Delta \varepsilon^{G}_{cv,\mbf{k'}(t")}\right].
\end{equation}

\section{Results}
\label{results}
\subsection{Isotropic band}
\begin{figure}
\includegraphics[width=85mm]{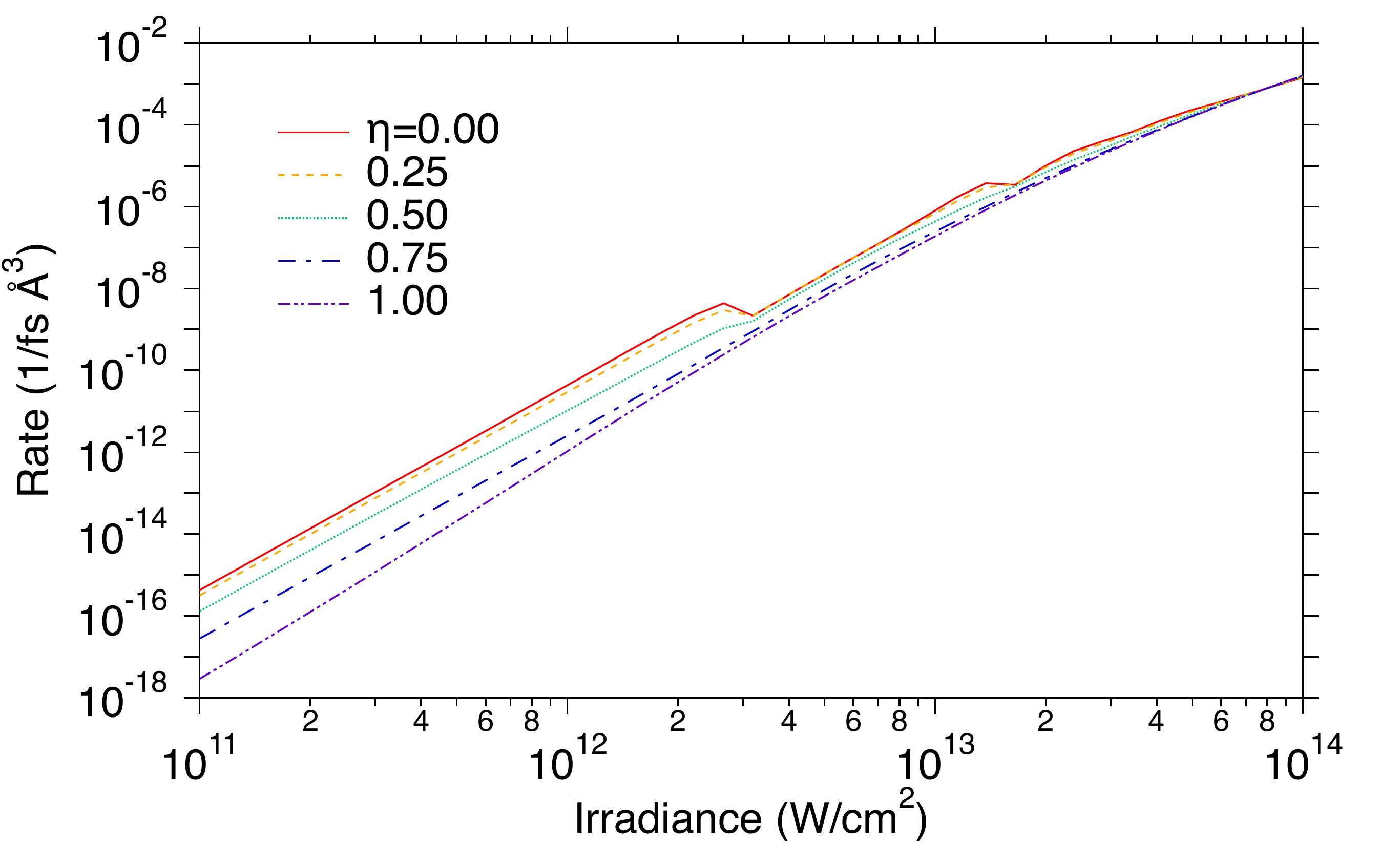}
\caption{\label{Dia_InDp} Laser irradiance dependence of the excitation rate with some ellipticity ($\eta$).}
\end{figure}

Diamond is a typical dielectric in non-linear laser-matter interaction studies, and we   
select it as an example to illustrate the application of the preceding formalism.
The transition probability of diamond by 
800~nm (1.55~eV) light is shown in Fig.~\ref{Dia_InDp}.
We assume a bandgap ($B_g$) of 7.4~eV and a reduced mass of 0.5~$m$.
The lowest order of the multi-photon absorption is the 5-photon process.

The excitation rates with linear (solid red line) 
and circular polarization (purple double-doted and dashed line)
are same as the previous results\cite{Kozak-2019}.
The excitation rate decrease as $\eta$ increases, which is consistent 
with the experimental results qualitatively.

\begin{figure}
\includegraphics[width=85mm]{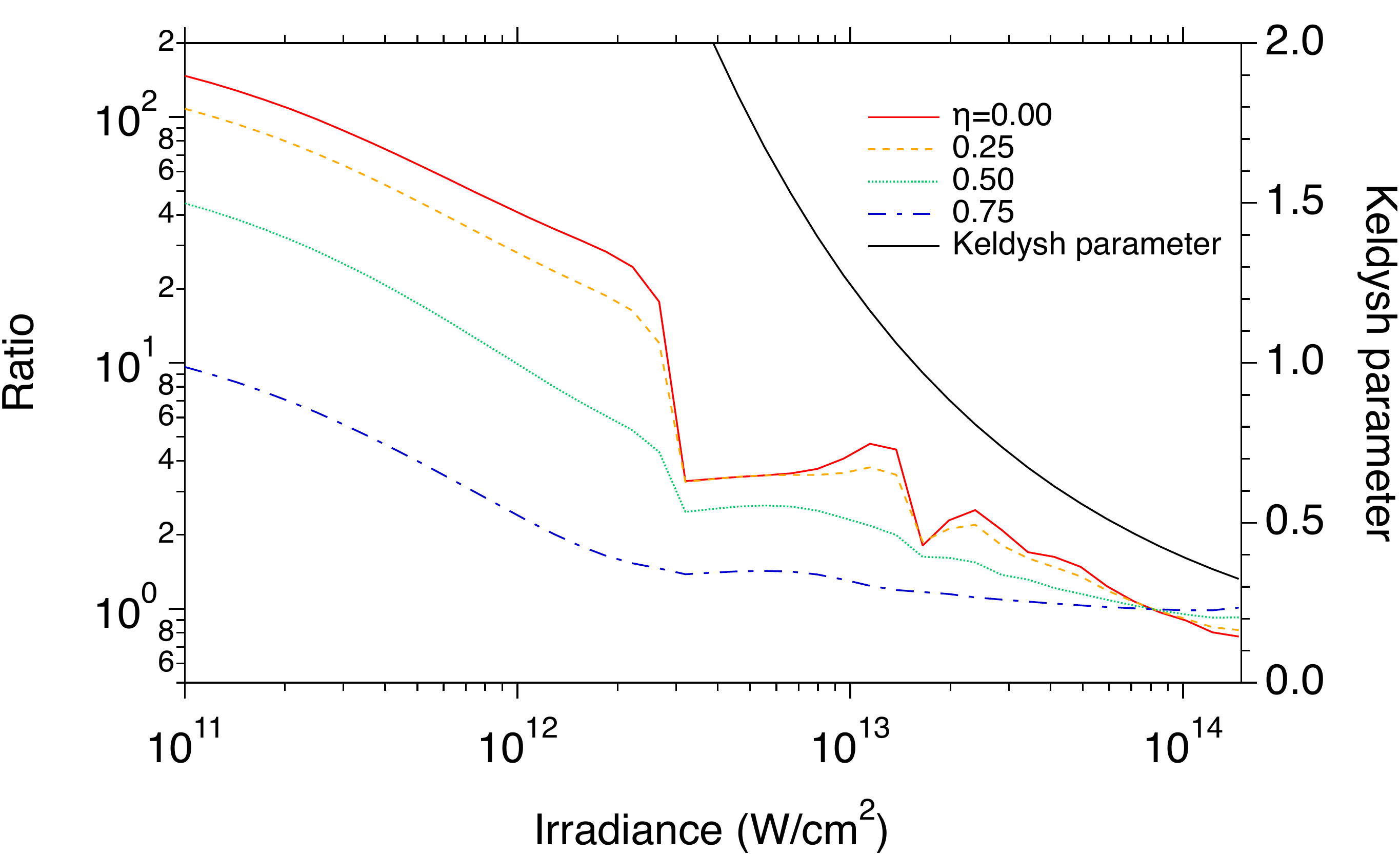}
\caption{\label{Ratio} Relative excitation rate with respect to  circular polarization.}
\end{figure}

The ratio of $W$ with respect to  circular polarization is shown in Fig.~\ref{Ratio}.
The ratio of $W$ accesses to one up to $7\times 10^{13}$ W/cm$^2$ for all $\eta$, while 
the ratio becomes less than one above $7\times 10^{13}$ W/cm$^2$.
To clarify the excitation process, we plot the Keldysh parameter ($\gamma_{k}$) in  Fig.~\ref{Ratio}
as a solid black line.
In general, the tunnelling process is dominant at $\gamma_{k} \ll 1.0$.
The reversing relationship in the ratio occurs at $\gamma_k < 0.5$, which suggests 
that the ellipticity dependence is different for the multi-photon and tunneling excitation processes. 

\begin{figure}
\includegraphics[width=85mm]{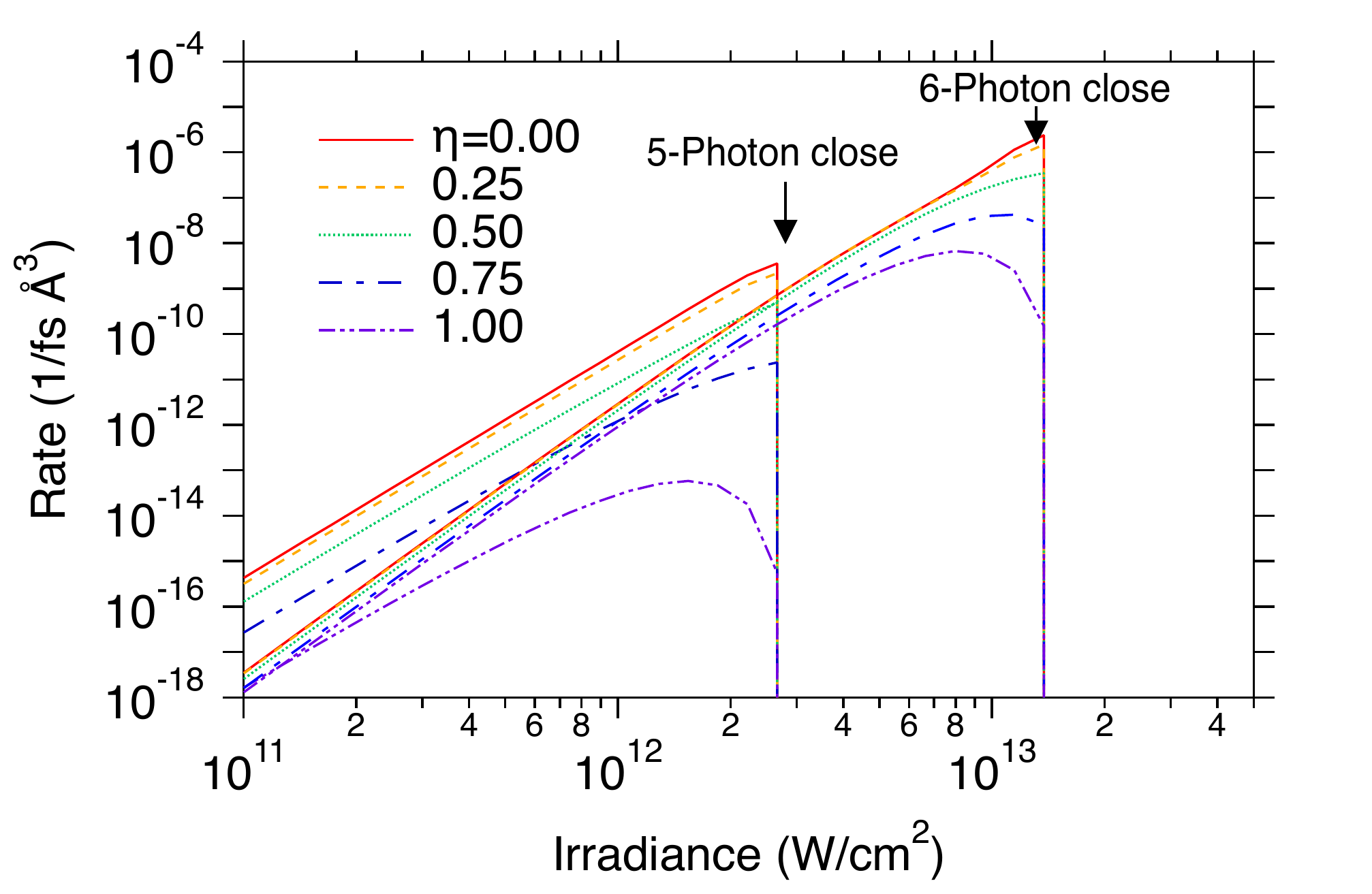}
\caption{\label{Dia_P_P} Excitation rate via 5- and 6-photon processes. The abrupt drop of the rate indicates the 
channel closing of each photonic process.}
\end{figure}

Although linear polarization shows clear 
5-photon absorption 
below $1\times 10^{12}$~W/cm$^{-2}$, 
the irradiance dependence of elliptical polarization
does not.
Figure~\ref{Dia_P_P} shows 5--photon absorption (5PA) and 
6--photon absorption (6PA), which is the $L=5$ and $6$ terms of 
Eq.~(\ref{eq:W_fin}) respectively.
The 5PA channel is closed at $2.6\times 10^{12}$~W/cm$^{-2}$, 
while the 6PA channel is closed at $1.4\times 10^{13}$~W/cm$^{-2}$.
The 5PA rate is larger than 
the 6PA absorption rate with
$\eta=0.0$ and $0.25$. 
The relationship between 5PA and 6PA is reversed as $\eta$ increases because the 5PA rate decreases significantly below the channel closing with higher $\eta$.
We have to consider some multi-photon processes to evaluate 
the excitation rate with elliptical and circular polarization.


\subsection{Anisotropic band}
In the previous subsection, we studied the $\eta$
dependence with isotropic parabolic band.
However, the $\eta$ dependence is much larger than 
the experimental result\cite{Kozak-2019}.
In this subsection, we would like to study how the 
band structure affects $\eta$ dependence.
Since the analytical formulation with 
direction-dependent reduced mass like Eq.~(\ref{mu_t_p}) is complicated, we solve Eq.~(\ref{TD_Ccv}) by real-time method.

\begin{figure}
\includegraphics[width=85mm]{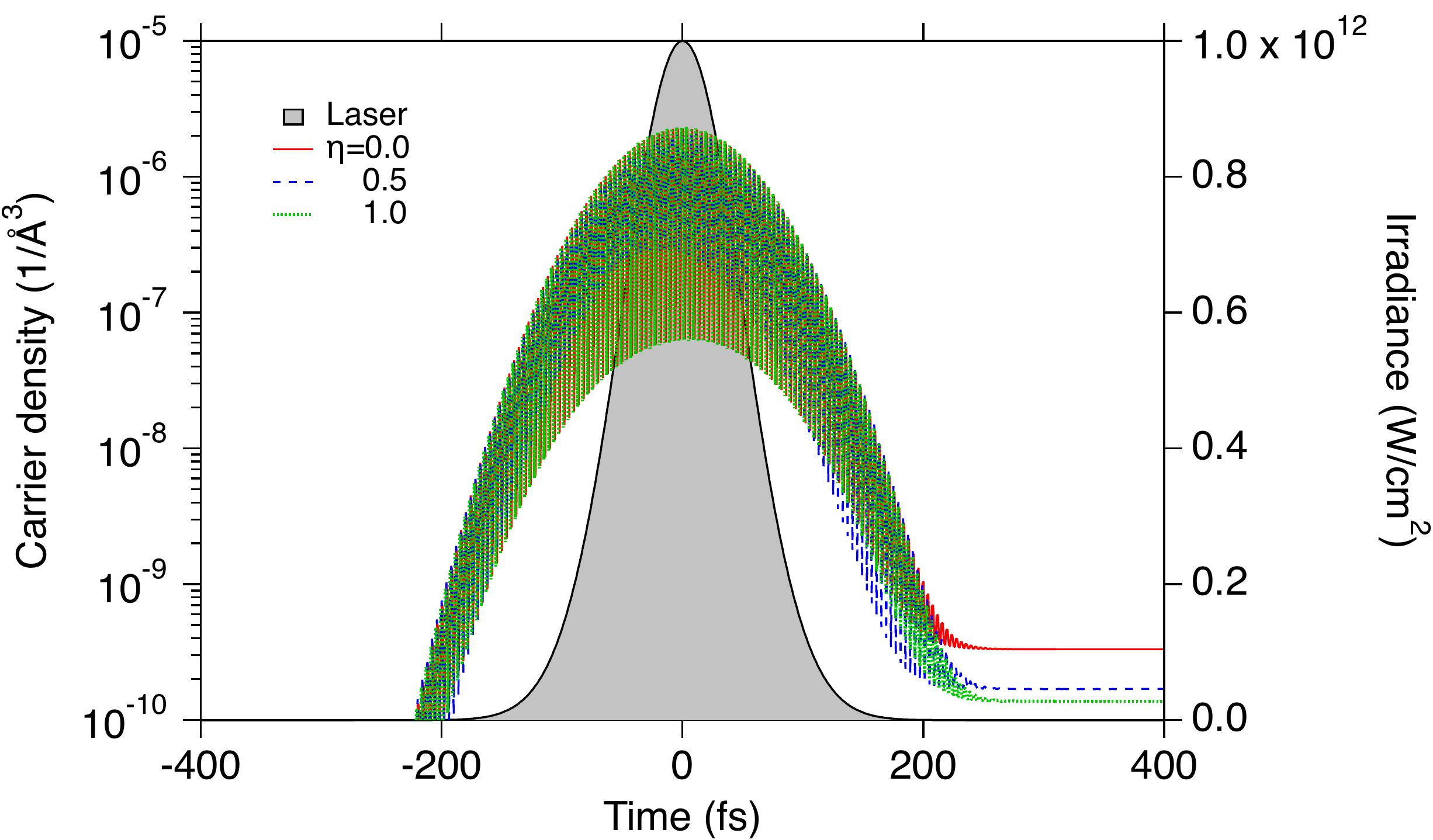}
\caption{\label{TD_Dens}Time evolution of carrier density with $\eta$ as 0, 0.5, and 1.0.
The grey shaded region is laser pulse irradiance. $\xi$ is fixed at $5\times 10^{-3}$. The solid line with gray shaded area indicates the time-evolution of the laser irradiance.}
\end{figure}

We assume the Gaussian-shape laser pulse,
\begin{equation}
    \mbf{A}(t)=A_0e^{-\frac{t^2}{\tau^2}}(\eta\sin\omega t, 0, \cos\omega t).
\end{equation}
Since diamond has 4--fold symmetry, the reduced mass can be mimicked by 
\begin{equation}
\label{mu_t_p2}
    \mu(\theta,\phi)=\mu_0\left[1+\xi \left(\sin^2 2\theta +\sin^2 2\phi\right) \right],
\end{equation}
where $\xi$ is the parameter for anisotropy of band structure.
Figure~\ref{TD_Dens} shows the time-evolution of the 
carrier density, 
\begin{equation}
    n(t)=\int d\mbf{k}|C^{\mbf{k}}_{vc}(t)|^2,
\end{equation}
with the ellipticity of 0.0, 0.5, and 1.0.
The irradiance is $1\times10^{12}$~W/cm$^2$ with frequency of 1.55~eV. The pulse duration parameter $\tau$ is 100~fs. 
The band structure parameter, $\xi$, is 0.005.
The isotropic band case indicates that the excitation rate with 
linear polarization ($\eta=0.0$) is about 50 times higher than that with circular polarization. 
On the other hand, the anisotropic case indicates a factor of 2.5, which is close to the factor of 5 in experiment\cite{Kozak-2019}.

\begin{figure}
\includegraphics[width=85mm]{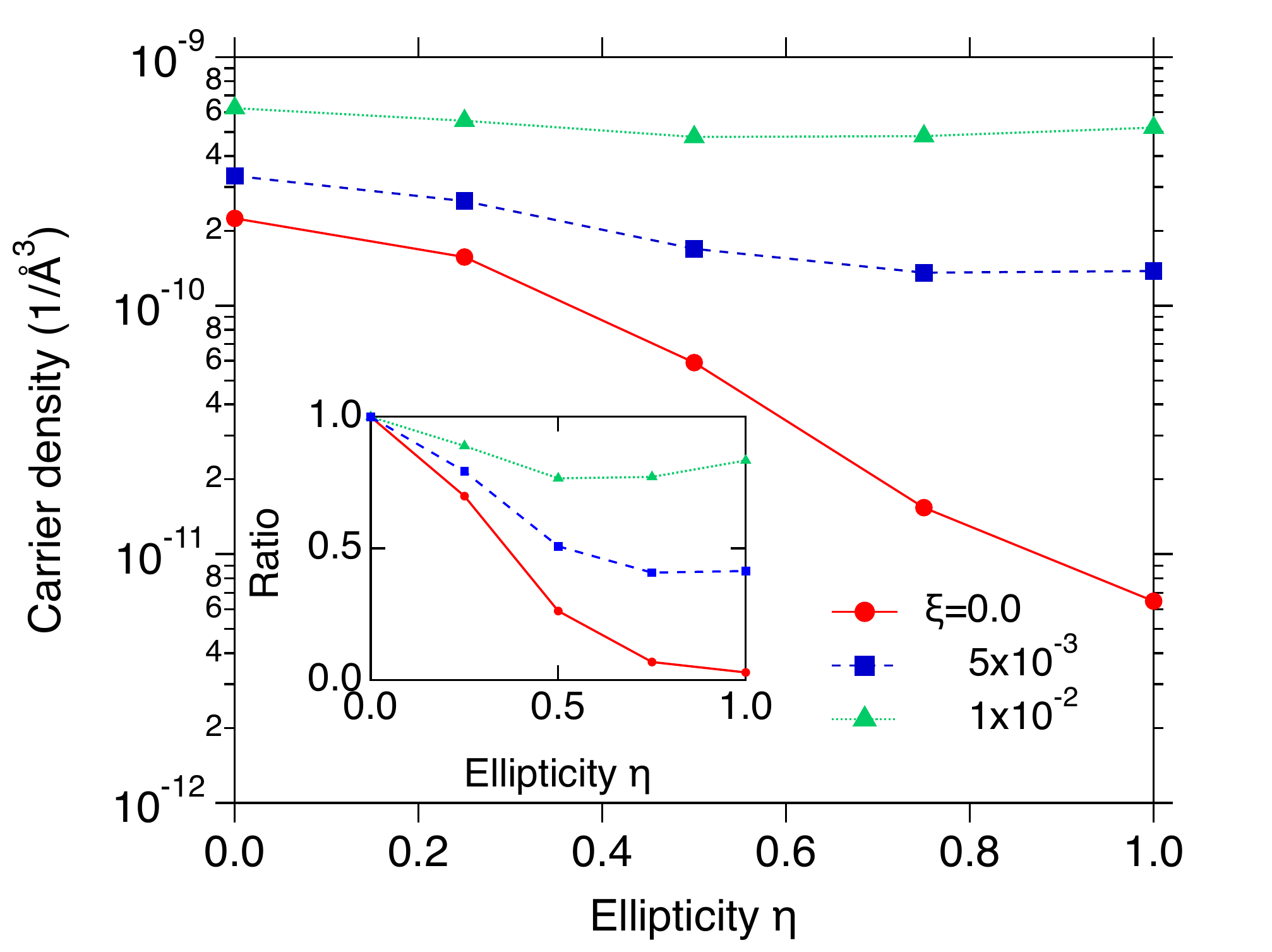}
\caption{\label{Ellip_Xi} Ellipticity and $\xi$ dependence of the carrier density at the end of time evolution. 
The inset figure shows the carrier density scaled at $\eta=0.0$.}
\end{figure}

Figure~\ref{Ellip_Xi} shows the $\xi$ and $\eta$ dependence of the carrier density.
The carrier density increases at higher $\xi$ in all $\eta$, and the $\eta$ dependence decreases as the $\xi$ increases.
In particular, in the case of $\xi=1\times 10^{-2}$, 
the excitation rate has the minimum at $\eta=0.5$.
The change of $\eta$ giving minimum excitation rate 
indicates that the band structure changes the $\eta$ dependence 
not only quantitatively but also qualitatively.

To clarify why the excitation $\xi$-dependence is 
more intense with higher ellipticity, we would like to
revisit the phase by the time integral of the energy.
If we see the electron at $\Gamma$-point with $\eta=1.0$, the 
time evolution of the relative energy is 
\begin{equation}
    \Delta \epsilon^{G}_{cv,\mbf{k'}(t)}
    =\frac{e^2 A_0^2}{\hbar^2 c^2\mu_0}\frac{1}{1+\xi\frac{1-\cos4\omega t}{2}}.
\end{equation}
The relative energy oscillates with a frequency of $4\omega$. On the other hand, in the linear polarization case, the relative energy oscillates with $2\omega$.
The higher relative energy oscillation frequency induces the increase of excitation rate.
Therefore, higher $\eta$ is more sensitive to 
$\xi$. 
Since the intra-band oscillation under elliptically polarized laser changes the $\eta$-dependence, the material with a significant anisotropic 
band structure may have the maximum excitation rate under circularly polarized laser.

It should be noted that $\xi$ affects excitation rate 
more than we expected from the experiment \cite{Kozak-2019}.
The sensitive behavior of our model may be due to the naive 
treatment of the band structure.
Diamond has almost isotropic band structure around the 
$\Gamma$-point and the direction dependence of effective 
mass becomes more prominent as $k_0$ increases.
Our results indicate that the realistic band structure 
is essential to reproduce the experiments quantitatively.

\section{Summary}
\label{conclusion}
We extended our previous analytical formula for the electron excitation rate in 
the dielectrics\cite{Otobe-2019} to elliptical polarization.
In general, ellipticity decreases the excitation rate.
In particular, the ellipticity dependence is significant in the isotropic system.
However, the decrease of excitation rate by elliptically polarized laser 
becomes small by assuming the anisotropic band structure because 
the symmetry-dependent oscillation of energy  
in the intra-band dynamics occurs with higher ellipticity.
If the anisotropy of the band structure is significant, 
the ellipticity increases the excitation rate.

The conflict between two experimental results reported by Temnov \textit{et. al.} \cite{Temnov-06}
and Koz\'ak \textit{et. al.}\cite{Kozak-2019}, 
and our numerical results
indicate that the anisotropy of the band structures 
is important to understand the laser excitation process.
\section{Acknowledgment}
\noindent  This research is supported by MEXT Quantum Leap Flagship Program (MEXT Q-LEAP)
under Grant No.  JPMX0118067246. 
This research is also partially supported by JST-CREST under Grant No. JP-MJCR16N5.
The numerical
calculations are carried out using the computer facilities of the
SGI8600 at Japan Atomic Energy Agency (JAEA).

\end{document}